\documentclass[]{article}
\usepackage{authblk}
\usepackage{cite}
\usepackage{bm}
\usepackage{pdfpages}
\usepackage{xcolor}
\usepackage{amsmath}

\begin{document}
	
\title{Elliptically polarized laser-pumped $M_x$ magnetometer towards applications at room temperature}

\author[1]{Shengran Su}
\author[1]{Guoyi Zhang}
\author[1]{Xin Bi}
\author[1]{Xiang He}
\author[1]{Wenqiang Zheng \thanks{Corresponding author: wqzheng@zjut.edu.cn}}
\author[1]{Qiang Lin \thanks{Corresponding author: qlin@zjut.edu.cn}}
\affil[1]{Collaborative Collaborative Innovation Center for Bio-Med Physics Information Technology, College of Science, Zhejiang University of Technology, Hangzhou 310023, China}

\maketitle

\begin{abstract}
	An atomic magnetometer operated with elliptically polarized light is investigated theoretically and experimentally. To explore the potential of this magnetometric configuration,  the analytical form of the outgoing signal is derived. Parameters that significantly influence the performance are optimized, which lead to a sensitivity of 300 $\rm fT/\sqrt{Hz}$ at 45 $^{\circ}$C with a 2$\times$2$\times2$ cm uncoated Rb vapor cell. It is remarkable that a sensitivity of 690 $\rm fT/\sqrt{Hz}$ is achieved at room temperature of 24 $^{\circ}$C, which is improved by an order of magnitude comparing with the conventional $M_x$ magnetometer under its own optimized condition. The elliptically polarized approach offers attractive features for developing compact, low-power magnetometers, which are available without heating the uncoated vapor cell.
\end{abstract}

\section{Introduction}
Optically pumped atomic magnetometers (AMs) have proved an immensely productive area of research, characterized by laser-based atomic techniques \cite{Budker07}. Sensitivity of AMs has approached and even exceeded superconducting quantum interference devices (SQUID) in the laboratory, without the requirement of cryogenics \cite{Dang10,Sheng13}. In addition to high sensitivity, other specific characteristics of magnetic detection devices are often required in different application scenarios, such as compact and wearable in biomagnetic monitoring \cite{Lembke14,Borna17,Zhang19,Boto18} and low power consumption in outdoor abnormal magnetic field detection \cite{Alexandrov92,Xu08} or space magnetic field measurement \cite{Zhang05}. 

In this paper, we discuss and investigate an AM operated with a single elliptically polarized beam, which can be identified as an elliptically polarized laser-pumped $M_x$ atomic magnetometer (EPMx AM).  This kind of configuration transplants the advantage of optical rotation (OR) detection mode to the conventional $M_x$ AM, where a circularly polarized pump beam is tuned to be resonant with $\rm D_1$ transition of the alkali atoms to simultaneously polarize the atomic spins and measure the spin polarization in the optical absorption (OA) detection mode \cite{Bloom62,Bison05}. It is a common sense that the OA mode facilitates a compact configuration of the sensor probe, since no extra beam is required to convey the atomic polarization information. The OR mode measures the atom-induced Faraday rotation angle of an additional off-resonance linearly polarized probe beam by the balanced polarimetry technique. The commonly used balanced polarimetry method involves a polarizing beam splitter (PBS) whose axis is rotated by $\pi/4$ with respect to the axis of a front linear polarizer. Then the two split beams severally fall onto photodetectors with differential outputs. Therefore the OR mode can suppress common mode noise and possess higher signal to noise ratio \cite{Miao19}. Treated as a combination of circular component and linear component, the elliptically polarized laser has been introduced to realize a compact atomic magnetometer working in the spin-exchange-relaxation-free (SERF) regime \cite{Shah09}. 
Circular component of the light creates relatively uniform spin polarization while the linear component is used to measure optical
rotation generated by the atoms. The SERF AM allows for femtotesla sensitivity. However, the SERF condition strongly depends on well magnetic shielding environment (below 10 nT) and high temperature atomic vapor (usually over 150 $^{\circ}$C).  
The limitations exclude SERF AMs for the use in magnetically unshielded environments and low-power scenarios. EPMx AMs preserve the compact potential of the single beam configuration, while promise a higher sensitivity than circularly polarized $M_x$ atomic magnetometer (CPMx AM). However, the condition of near-resonant light-atom interaction in EPMx AMs leads to a different optimization process, comparing with SERF AM using far off-resonance elliptically polarized light. 

Here we theoretically and experimentally optimize the parameters of EPMx AM, including frequency, ellipticity and intensity of the laser. Finally the EPMx AM shows significant superiority in sensitivity, compared with the optimized $M_x$ AM using OA detection mode. A sensitivity of 300 $\rm fT/\sqrt{Hz}$ at 45 $^{\circ}$C is achieved
with a 2$\times$2$\times2$ cm uncoated Rb vapor cell.
The outperformance is particularly remarkable in the room temperature. At 24 $^{\circ}$C, the sensitivity is improved from 7.57 $\rm pT/\sqrt{Hz}$ to 0.69 $\rm pT/\sqrt{Hz}$ by introducing elliptically polarized laser.
Our experimental results also show that the sensitivity of EPMx AM is less dependent on temperature than its conventional counterpart.   

\section{Principle of EPMx AM}
\subsection{Basic configuration and experimental setup of EPMx AM}

\begin{figure}[htp]
	\centering\includegraphics[width=8cm]{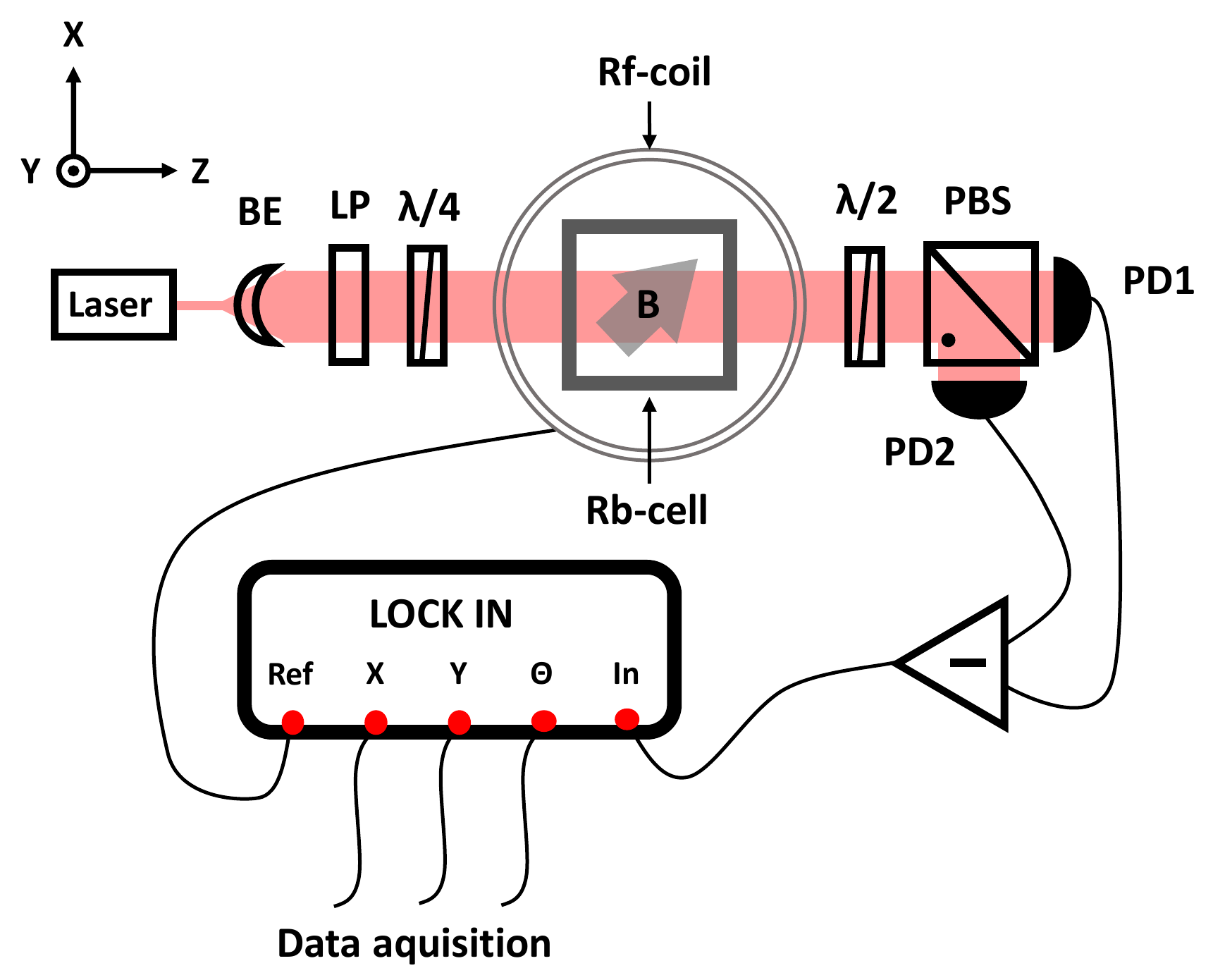}
	\caption{{Schematic diagram of the experimental setup. BE: beam expander, LP: linear polarizer, $\lambda$/4: quarter-wave plate, $\lambda$/2: half-wave plate, PBS: polarized beam splitter, PD: photodiode.}}
	\label{setup}
\end{figure}

The optical system to realize an EPMx AM is shown in Fig. \ref{setup}, representing the experimental arrangement used in this work. A cubic cell with a volume of 8 $\mathrm{cm^3}$ contains a drop of enriched $^{87}$Rb atoms and 200 Torr of $\mathrm{N_2}$ gas for quenching and slowing atomic diffusion. The cell is placed inside a five-layer magnetic shield cylinder, where three pairs of orthogonal internal coils can generate a stable, well-defined magnetic bias field in the xz-plane and an oscillating radio frequency (rf) magnetic field along the y-axis. 
The cell could be heated in a nonmagnetic oven by high-frequency ac currents at 70 kHz and its temperature is stabilized by a closed loop. An extended-cavity diode laser along the z-axis, tuned close to $^{87}$Rb $\rm D_1$ transition, was used in our experiment. Before illuminating the cell, the light passed through a linear polarizer and a quarter-waveplate with its optic axis oriented at an angle $\phi$ relative to the linear polarizer. The ellipticity of the light could be adjusted by changing the angle $\phi$. In this configuration, the electric feld of the light injected into the cell could be represented as a superposition of left-circular ($\sigma^+$, $ \bm{{\cal L}} = {{e^{i2\pi \nu  t}}/\sqrt 2 }$) and right-circular ($\sigma^-$, $\bm{{\cal R}} = {{e^{ - i2\pi \nu t}/\sqrt 2}}$) basis components, in the form
\begin{equation}
{E_{in}}\left( {z = 0} \right) = {E_0}\left( {\frac{{\cos \phi  + \sin \phi }}{{\sqrt 2 }}\bm{{\cal L}} + \frac{{\cos \phi  - \sin \phi }}{{\sqrt 2 }}\bm{{\cal R}}} \right),
\label{injectlight}
\end{equation}
where $\nu$ is the laser frequency. The probabilities of a photon being in the $\sigma^+$ and $\sigma^-$ states are $\left( {1 + s} \right)/2$  and $\left( {1 - s} \right)/2$, respectively.  $s= \sin (2 \phi)$ characterizes the average photon spin component along z direction. It ranges from -1 to +1, where $s=-1$ corresponds to $\sigma ^-$ light, $s=0$ corresponds to linearly polarized light, and $s=+1$ corresponds to $\sigma ^+$ light. The optical rotation from atoms was subsequently converted to an electric signal through a balanced polarimeter, consisting of a half-wave plate, a PBS and a balanced photodetector. The half-wave plate was rotated to balance the intensity in the balanced photodetector in the absence of optical rotation from atoms. The optoelectronic signal was fed to the input of a lock-in amplifier. The digital lock-in amplifier demodulated the oscillating signal from optical rotation with reference to the applied oscillating rf magnetic field. 

In our experiment, the diameters of the laser beam is 6 mm. The internal side length of the cell is 17 mm. Plenty of $\mathrm{N_2}$ gas (200 Torr) make the diffusion of $^{87}$Rb atoms negligible. The cross-talk free distance of $^{87}$Rb atoms are about 0.8 mm at 100 $^{\circ}$C \cite{Kim14}. Therefore, the whole sensing zone is actually a cylinder with a diameter of laser beam and a length of the internal cell. The sensing volume is about 0.48 $\mathrm{cm^3}$.

\subsection{Laser-pumped $M_x$ magnetometer}

The $M_x$ magnetometer is sensitive to the modulus of the external magnetic field by measuring the Larmor frequency of atoms. A static magnetic field $\bm {\mathrm B_0}$ is aligned in xz plane. Magnetic resonance technique is 
introduced by employing an oscillating magnetic field ${\bm {\mathrm {B_{rf}}}} = 2{B_{rf}}\cos {\omega _{rf}}\hat y$ perpendicular to $\bm {\mathrm B_0}$ and the propagation direction of light. The amplitude $B_{rf}$ is much smaller than $B_0$. The overall evolution of the atomic spin angular momentum $\bm {\mathrm S}$ is well-described by the Bloch equation
\begin{equation}
\frac{d \bm {\mathrm S}}{\rm{dt}}=\gamma \cdot \bm {\mathrm S}\times \bm {\mathrm B}+\Gamma _P\cdot \left(\bm {\mathrm {S_0}}-\bm {\mathrm S}\right)-{\Gamma_{rel}} \cdot \bm {\mathrm S},
\label{bloch}
\end{equation}
Where $\gamma=7$ Hz/nT is the gyromagnetic ratio of $^{87}$Rb atomic spins, $\bm {\mathrm B} = \bm {\mathrm B_0} + \bm {\mathrm {B_{rf}}}$ is the total magnetic field, $\Gamma _P$ is the pumping rate,
${\bm {\mathrm {S_0}}}=\hat zs{\Gamma _p}/[2\left( {{\Gamma _p} + {\Gamma _{rel}}} \right)]$ is the equilibrium atomic spin angular momentum in the absence of the oscillating excitation and $\Gamma_{rel}$ is the spin-relaxation rate. We further define $\Gamma_{rel}^1$ ($\Gamma_{rel}^2$) as the longitudinal (transverse) spin-relaxation rate.
The fist term of the Bloch equation describes the precession of $\bm {\mathrm S}$ around the magnetic field $\bm {\mathrm B}$, the second term represents the optical pumping process,
and the third term describes the spin relaxation process. The projection of the precessing polarization onto the propagation direction of the light beam then leads to an oscillating polarization component along that axis, and therefore to a periodic modulation of the optical rotation. Through the steady-state solution to Eq.\ref{bloch} in the rotating frame with angular frequency $\omega _{rf}$, we can obtain the quadrature amplitude $P_{\rm{qu}}$ and in-phase amplitude $P_{\rm{ip}}$ of the photocurrent with respect to the oscillating magnetic field, which are given by
\begin{equation}
{P_{{\rm{qu}}}}\left( \delta  \right) = {P_0}\sin \left( {2\vartheta } \right)\frac{{\Omega {\Gamma _2}}}{{{\Omega ^2}{\Gamma _2}/{\Gamma _1} + {\Gamma _2}^2 + {\delta ^2}}}.
\label{pqu}
\end{equation}
\begin{equation}
{P_{{\rm{ip}}}}\left( \delta  \right) = {P_0}\sin \left( {2\vartheta } \right)\frac{{\delta \Omega }}{{{\Omega ^2}{\Gamma _2}/{\Gamma _1} + {\Gamma _2}^2 + {\delta ^2}}} 
,\label{pip}
\end{equation}
Here $\vartheta$ is the angle between the direction of laser beam and $\bm {\mathrm B_0}$, $\Omega$ is the Rabi frequency, \begin{math}\delta=\omega_{rf}-\gamma B_0\end{math} is the detuning of the oscillating field $\bm {\mathrm {B_{rf}}}$ from the Larmor frequency, ${\Gamma _{1(2)}} = \Gamma _{rel}^{1(2)} + {\Gamma _P}$ is the effective longitudinal (transverse) polarization-relaxation rate.  $P_0$ is the equilibrium atomic polarization and is defined as 
\begin{equation}
{P_0} = 2\left\langle {{S_0}} \right\rangle  = \frac{{s{\Gamma _p}}}{{{\Gamma _p} + {\Gamma _{rel}}}}.
\label{P0}
\end{equation}
We set $\vartheta=45^ \circ$ to maximize the signal amplitude. We can see that $P_{\rm{qu}}$ has an absorptive Lorentzian line shape, and $P_{\rm{ip}}$ has a dispersive Lorentzian line shape with the same half-width expressed as
\begin{equation}
\Delta {\omega _{{\rm{HW}}}} = \sqrt {{\Omega ^2}{\Gamma _2}/{\Gamma _1} + {\Gamma _2}^2} .
\end{equation}
The phase $\theta$ between $P_{\rm{ip}}$ and $P_{\rm{qu}}$ can be calculated as
\begin{equation}
\tan \left( \theta  \right) = \frac{{{P_{{\rm{ip}}}}}}{{{P_{{\rm{qu}}}}}} = \frac{{{\delta}}}{\Gamma _2}.
\label{theta}
\end{equation}
The width of the phase signal is smaller than $\Delta {\omega _{\rm{HW}}}$ since it is immune to rf power broadening and can be expressed as
\begin{equation}
\Delta \omega _{\rm{HW}}^\theta  = \Gamma _2.
\label{halfwidth}
\end{equation}

\begin{figure}[h!]
	\centering
	\includegraphics[width=8cm]{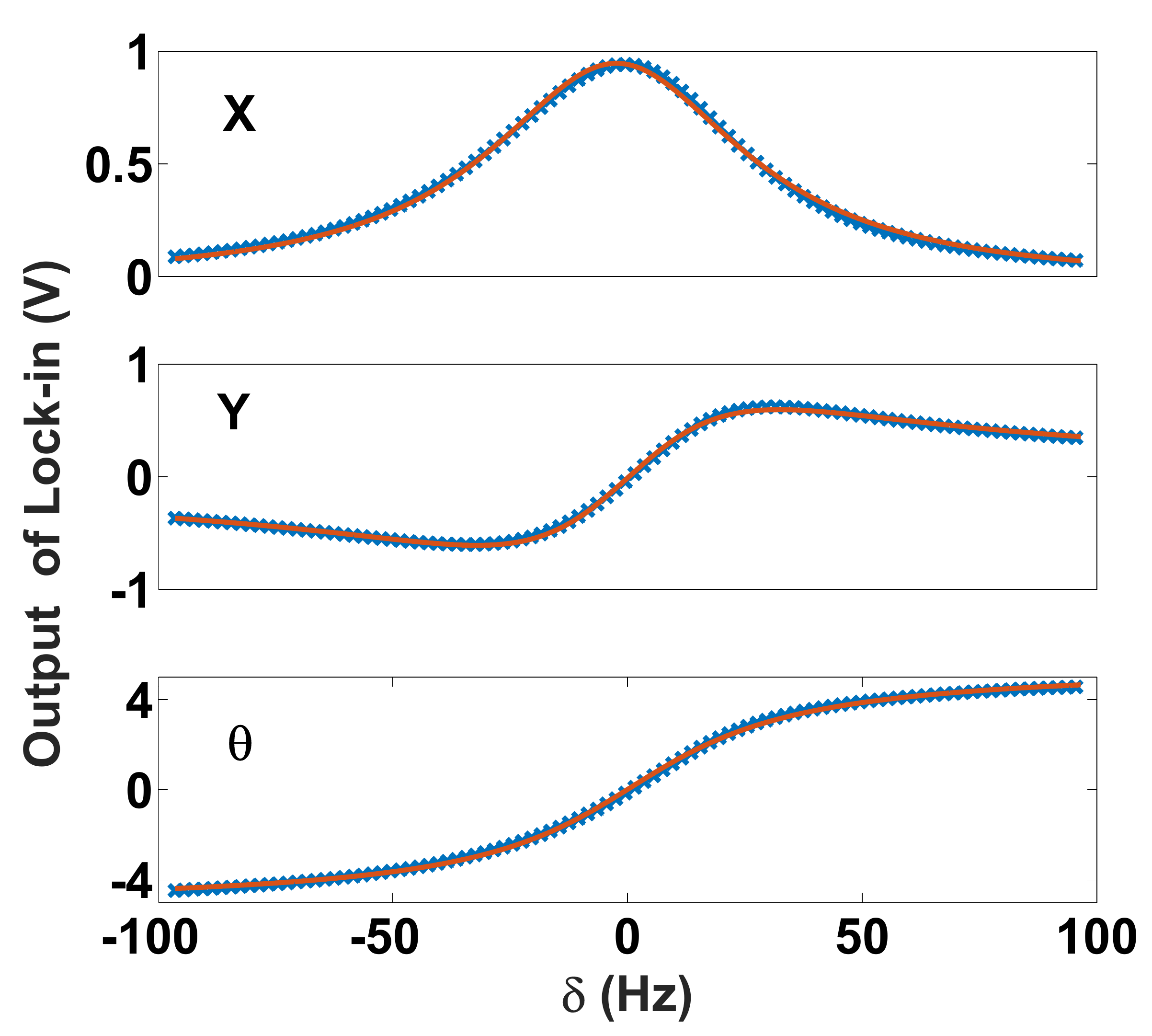}
	\caption{Measured magnetic-resonance line shapes of the 
		quadrature (top), in-phase (middle) and their relative phase (bottom) signals by scanning $B_{\rm{rf}}$ frequency. The blue dots represent the experimental data, while the red solid lines are their fitted Lorentzian line shapes. The fitted 
		half-widths are 32 Hz, 31 Hz and 26.7 Hz for X, Y and $\theta$ outputs, respectively.}
	\label{xytheta}
\end{figure}

In resonance the phase is $90^\circ$ and near resonance it has a linear dependence on the detuning $\delta$. Therefore the Larmor frequency and hence the magnetic field can be inferred by measuring the phase $\theta$.  
For a $B_{\rm{rf}}$ frequency scan at a fixed $B_0$,
the quadrature output X, in-phase output Y and phase output $\theta$ of the lock-in amplifier, as well as their fitted Lorentzian line shapes, are shown in Fig \ref{xytheta}.

\section{Mechanism analysis and optimization}
\subsection{Optical pumping}

The natural broadening due to limited lifetime of the excited state, pressure broadening due to collisions with buffer gas, and Doppler broadening due to atomic thermal velocity, are three main effects contributing to the form of atomic frequency response to photons. For the transition $F$ (ground state) $\to$ $F'$ (excited state), the photon absorption cross-section can be generally expressed by a voigt profile \cite{Happer67,Seltzer08}, as
\begin{equation}
{\sigma _{F,F'}}\left( \nu  \right) = \pi {r_e}c{f_{D1}}{\mathop{\rm Re}\nolimits} \left[ {{\cal V}\left( {\nu  - {\nu _{F,F'}}} \right)} \right],
\end{equation}
where $\nu_{F,F'}$ is the resonance frequency of the related transition, $r_e$ is the classical electron radius, $c$ is the speed of light and ${f_{D1}} \approx {1 \mathord{\left/{\vphantom {1 3}} \right. \kern-\nulldelimiterspace} 3}$ is the transition strength for alkali atoms. The voigt profile is calculated by 
\begin{equation}
\mathcal{V}\left(\nu -\nu _{F,F'}\right) =\int_0^{\infty }  \mathcal{L} \left(\nu -\nu '\right) \mathcal{G}  \left(\nu ' -\nu _{F,F'}\right)\, d\nu '.
\label{voigt}
\end{equation}
It is a comprehensive result of a Lorentzian curve $\mathcal{L}(\nu)$ with full width at half maximum(FWHM)  $\Gamma _L =\Gamma _{\rm{nature}}+\Gamma _{\rm{pressure}}$ and a Gaussian curve $\mathcal{G}(\nu)$ with FWHM  $\Gamma _{\rm{Doppler}}$. The Voigt profile has a explicit complex form \cite{Seltzer08}
\begin{equation}
\mathcal{V}\left(\nu -\nu _0\right) =\frac{2 \sqrt{{\ln 2}/{\pi }}}{\Gamma _G} w\left(\frac{2 \sqrt{{\ln 2}}\left(\nu -\nu _0\right)+{i \Gamma _L}/{2}}{\Gamma _G}\right),    
\label{wvoigt}
\end{equation}    
where function \begin{math}w\left(x\right)\end{math} is the complex error function, given by 
\begin{equation}
w\left(x\right)=e^{-x^2} (1-\rm{erf} (-i x)).
\label{w}
\end{equation}

\begin{figure}[htbp]
	\centering\includegraphics[width=8cm]{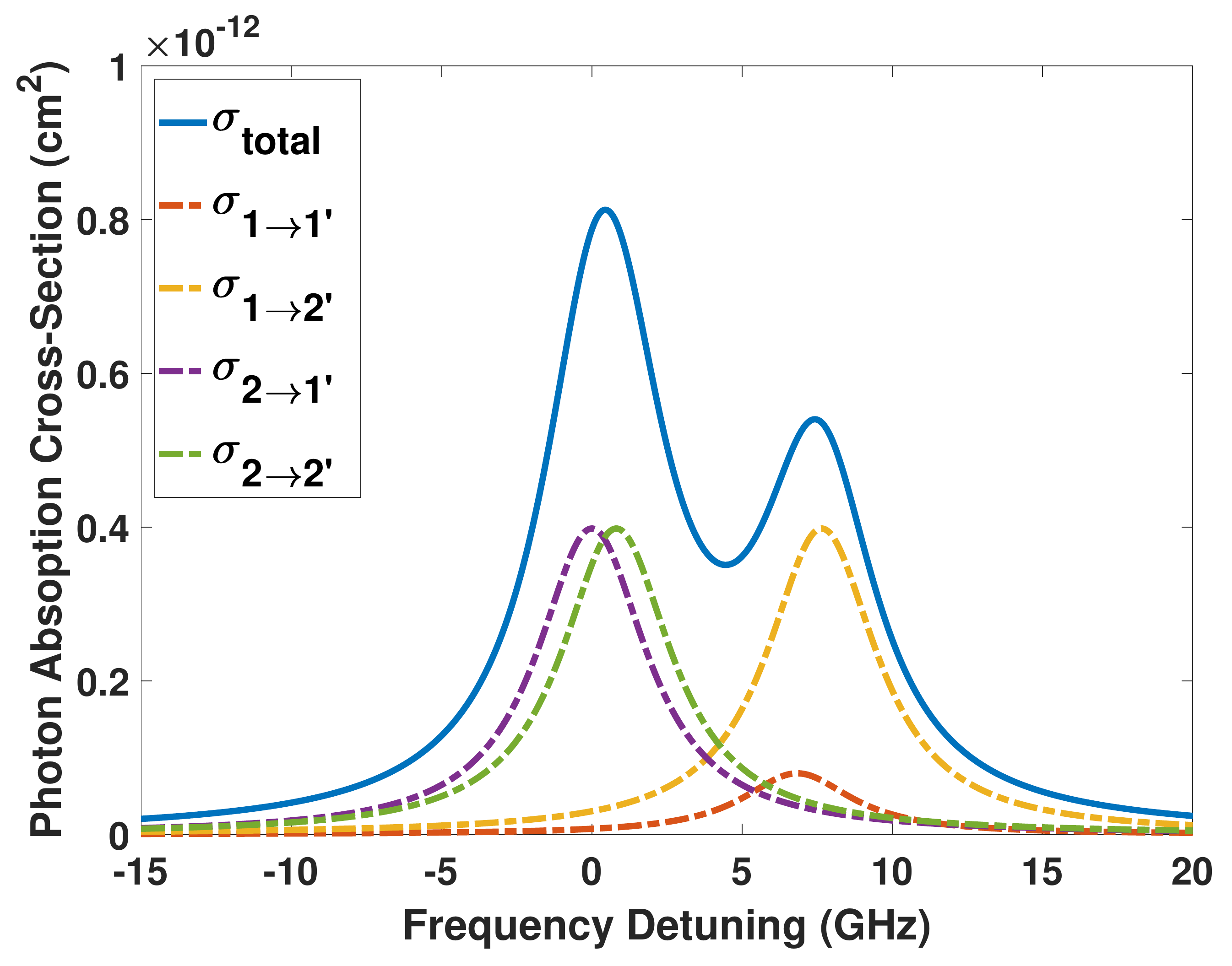}
	\caption{Atomic frequency response for optical absorption 
		near the $^{87}$Rb D$_1$ transition, taking into account the hyperfine splitting of the ground and excited states (Eq. (\ref{totalcrosssection})). Frequency is expressed as detuning from the transition $F=2$ $\to$ $F'=1$. Corresponding to our experimental condition, we set $\Gamma _G$=0.53 GHz (Doppler broading at 333$K$) and $\Gamma _L$=4.36 GHz (pressure broadening caused by 200 torr nitrogen gas).}
	\label{crosssection}
\end{figure}

Under our experimental condition, the Doppler broadening is about 0.53 GHz and pressure broadening is about 4.36 GHz. They are comparable to the hyperfine splitting of $^{87}$Rb atoms (6.84 GHz for the ground state ${}^2{S_{{1 \mathord{\left/{\vphantom {1 2}} \right. \kern-\nulldelimiterspace} 2}}}$, 0.81 GHz for the excited state ${}^2{P_{{1 \mathord{\left/{\vphantom {1 2}} \right. \kern-\nulldelimiterspace} 2}}}$). Therefore 
it is necessary to separately consider the four individual resonances in  D$_1$ transition. The total photon absorption cross-section is given by
\begin{equation}
\sigma _{\rm{total}}\left(\nu\right)=\pi r_ecf_{D1}\sum _{F,F'} A^{abs}_{F,F'} {\mathop{\rm Re}\nolimits} \left[ {{\cal V}\left( {\nu  - {\nu _{F,F'}}} \right)} \right],
\label{totalcrosssection}
\end{equation}
where $A^{abs}_{F,F'}$ is the normalized relative strength for the transition $F$ $\to$ $F'$. Their values for atoms with nuclear spin $I = 3/2$ are given in Tab. \ref{table}. The lineshape of $\sigma _{\rm{total}}$, which determines the optical absorption lineshape of $^{87}$Rb atoms near the D$_1$ transition, is shown in Fig. \ref{crosssection}.

\begin{table}[htbp]
	\centering\includegraphics[width=8cm]{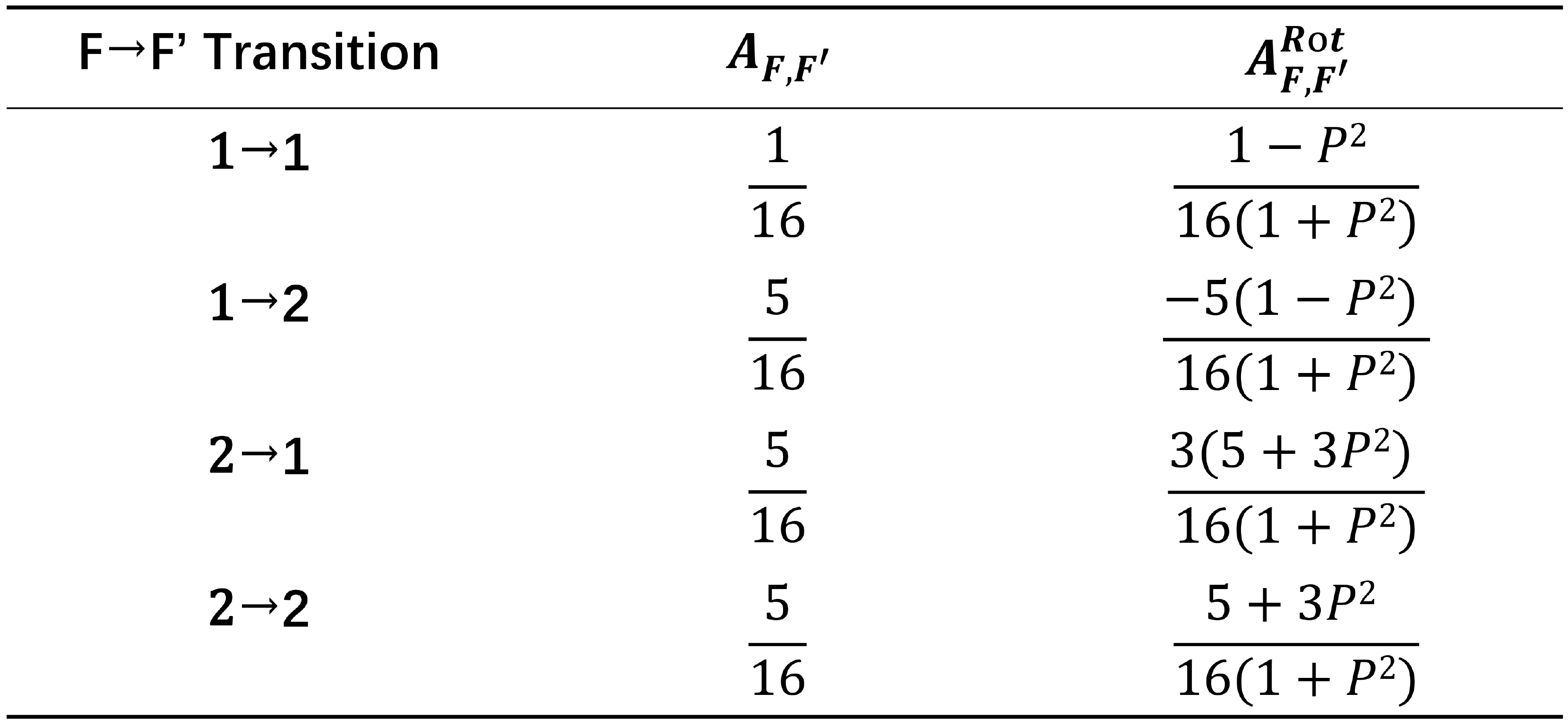}
	\caption{Relative strengths $A^{abs}_{F,F'}$ of the individual D$_1$ hyperfine resonances for photon absorption and relative strengths $A_{F,F'}^{rot}$ for optical rotation as functions of polarization $P$ \cite{Seltzer08}.}
	\label{table}
\end{table}

The pumping rate $\Gamma_P\left(\nu\right)$ at which an atom absorbs photons of frequency $\nu$ is
\begin{equation}
\Gamma_P\left(\nu\right)=\sigma _{\rm{total}}\left(\nu\right)\Phi\left(\nu\right),
\label{pumpingrate}
\end{equation}
where $\Phi\left(\nu\right)$ is the total flux of photons of frequency $\nu$ incident on the atom in units of number of photons per area per time. The equilibrium atomic polarization of Eq. (\ref{P0}) can be rewritten as 
\begin{equation}
{P_0} = \frac{{s{\sigma _{\rm{total}}}\left( \nu  \right)\Phi \left( \nu  \right)}}{{{\sigma _{\rm{total}}}\left( \nu  \right)\Phi \left( \nu  \right) + {\Gamma _{rel}}}}.
\label{pumpingrate}
\end{equation}

\subsection{Optical rotation}

\begin{figure}[t]
	\centering\includegraphics[width=8cm]{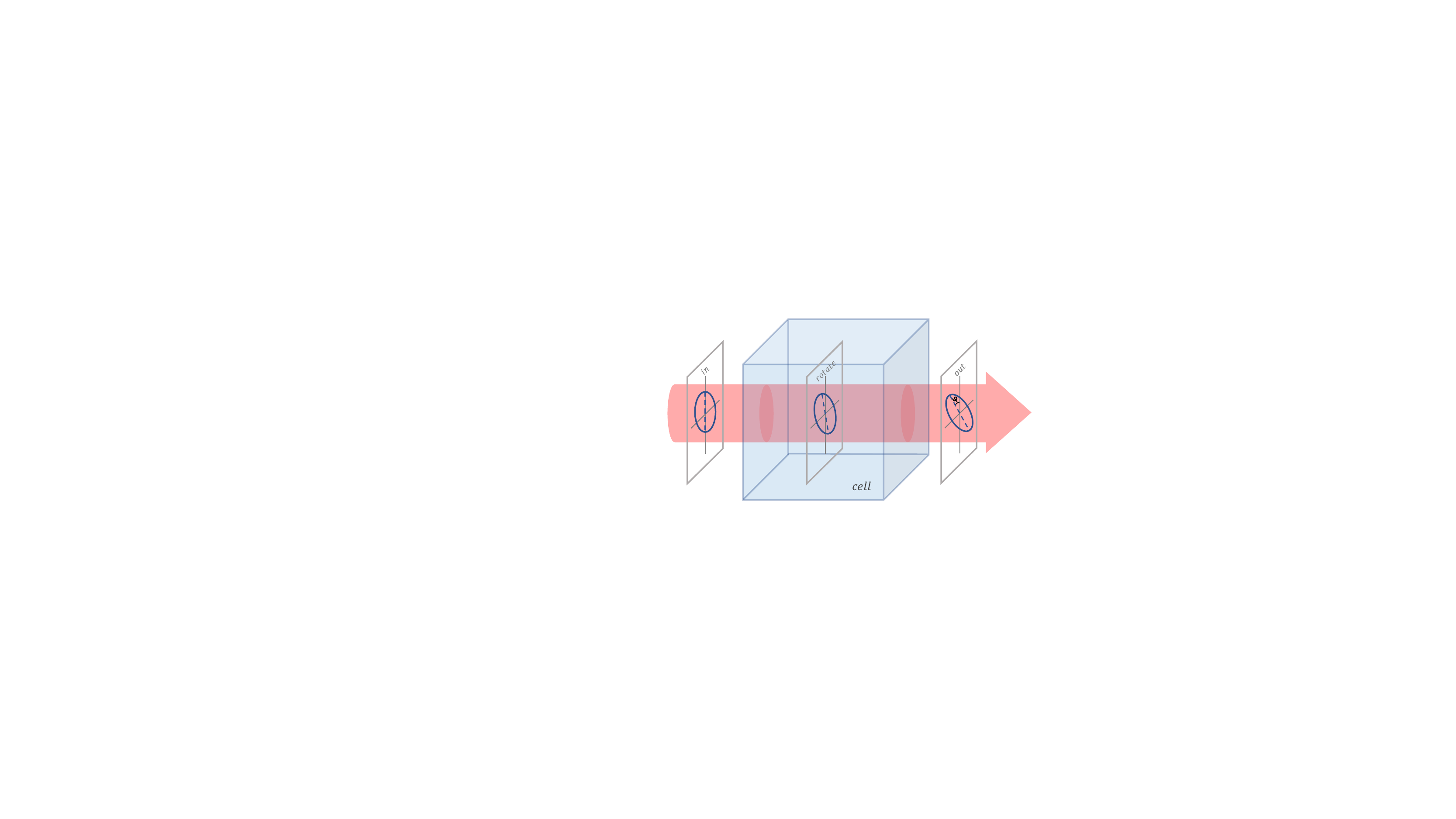}
	\caption{Optical rotation of an elliptically polarized light. The major axis of the polarization ellipse rotates an angle $\varphi$ when the light experiences optical rotation.}
	\label{rotation}
\end{figure}

Polarized atomic ensemble is a birefringent medium and  can induce
the phenomenon of optical rotation. That is, the polarization plane of the light rotates by an angle $\varphi$ when it passes through the vapor cell due to the
different refractive indices for $\sigma^+$ light and $\sigma^-$ light, which can be expressed as
\begin{equation}
{n_ + }\left( \nu  \right) = 1 + \frac{{n{r_e}{c^2}{f_{{\rm{D1}}}}}}{{2\nu }}\frac{{1 + {P}}}{2}\sum\limits_{F,F'} {A_{F,F'}^{{rot}}} Im\left[ {{\cal V}\left( {\nu  - {\nu _{F,F'}}} \right)} \right],
\label{nzheng}
\end{equation}
\begin{equation}
{n_ - }\left( \nu  \right) = 1 + \frac{{n{r_e}{c^2}{f_{{\rm{D1}}}}}}{{2\nu }}\frac{{1 - {P}}}{2}\sum\limits_{F,F'} {A_{F,F'}^{{rot}}} Im\left[ {{\cal V}\left( {\nu  - {\nu _{F,F'}}} \right)} \right],
\label{nfu}
\end{equation}
where $n$ is the atomic number density of $^{87}$Rb vapor, $P$ is the atomic polarization along the propagation direction of the light. The atomic polarization 
somewhat deviates from the propagation direction of the light under the continuous radiation of the radio-frequency field. In fact, we have taken this deviation into account during the derivation of the output in the $M_x$ configuration (Eq. \ref{pqu},  Eq. \ref{pip}), so we have $P = {P_{ip}}\cos \omega_{rf} t + {P_{qu}}\sin \omega_{rf} t$ with amplitude $A_P = \sqrt {{P_{qu}}^2 + {P_{ip}}^2}$. Hyperfine splittings are necessarily considered when we pursue an accurate quantitative description of optical rotation. Similar to $A_{F,F'}^{abs}$, $A_{F,F'}^{rot}$ are defined as the normalized relative strengths concerning the optical rotation process. $A_{F,F'}^{rot}$ are functions of the polarization $P$. We list their functional relations with $P$ for $^{87}$Rb D$_1$ transition in Tab. \ref{table}.
When the incident elliptically polarized light in the form of Eq. \ref{injectlight} passes through the vapor cell, the $\sigma^+$ and  $\sigma^-$ components separately accumulate phases under their own refractive indices. Assuming the length of the cell is $l$, the emergent light at the end of the cell can be written as
\begin{equation}
{E_{out}}\left( {z = l} \right) = {E_1}\left( {\frac{{\cos \phi_1 + \sin \phi_1}}{{\sqrt 2 }}{e^{i{\varphi_+}}}\bm{{\cal L}} + \frac{{\cos \phi_1 - \sin \phi_1}}{{\sqrt 2 }}{e^{i{\varphi_-}}}\bm{{\cal R}}} \right),
\label{Eout}
\end{equation}
where $\varphi_{+(-)}={2\pi \nu l{n_ {+(-)} }}/c$. Note that different symbols $E_1$ and $\phi_1$ are used here, comparing with Eq. (\ref{injectlight}). It indicates that changes take place in intensity and ellipticity after the near-resonant light passes through the cell due to optical 
absorption. We will discuss this topic in the next section. Ignoring the common phase, we get the optical rotation angle
\begin{equation}
\varphi = \frac{{{\varphi _ + } - {\varphi _ - }}}{2} =\frac{ \pi \nu l}{c}\left(n_+-n_-\right).
\label{rotationangle}
\end{equation}
As we know, in general cases the endpoint of the electric field vector traces out an ellipse in one cycle. Correspondingly, the optical rotation can be regarded as  
a rotation of the major axis of the ellipse, which is 
diagrammatically shown in Fig. \ref{rotation}. Then the optical rotation is measured by the balanced polarimeter shown in Fig. \ref{setup}. The differential signal $\cal D$ of the photodetector is given by
\begin{equation}
{\cal D} \propto {{E_1}}^{2}\sqrt {1 - {{s_1}^2}} \sin \left( {2\varphi } \right) \\
\propto {{\Phi_1}}\sqrt {1 - {{s_1}^2}} \sin \left( {2\varphi } \right),
\label{Dsignal}
\end{equation}
where $\Phi_1$ is the photon flux of the emergent light. When the magnetic resonance condition $\delta=\omega_{rf}-\gamma B_0=0$ is satisfied, the amplitude of the oscillating photoelectric signal is
\begin{equation}
\begin{split}
Sig\left( {s,\nu } \right) &\propto {{\Phi_1}}\sqrt {1 - {{s_1}^2}} \frac{{2\pi \nu l}}{c}\left( {{n_ + } - {n_ - }} \right)  \\
&\propto {{\Phi_1}}\sqrt {1 - {{s_1}^2}} P{C_{rot}} \\
&\propto {{\Phi_1}}\sqrt {1 - {{s_1}^2}} {P_0}\frac{{{\Gamma _2}}}{{{\Omega ^2}{\Gamma _2}/{\Gamma _1} + {\Gamma _2}^2}}{C_{rot}}   \\
&\propto s\sqrt {1 - {{s_1}^2}} \frac{{{{\Phi_1}}{\Phi _0}{\sigma _{{\rm{total}}}}}}{{{\sigma _{{\rm{total}}}}{\Phi_0} + {\Gamma _{rel}}}}\frac{{{\Gamma _2}}}{{{\Omega ^2}{\Gamma _2}/{\Gamma _1} + {\Gamma _2}^2}}{C_{rot}},
\end{split}
\label{sig}
\end{equation}
where ${{\Phi _0}}$ is the photon flux of the incident light and ${C_{rot}}{\rm{ = }}\sum\nolimits_{F,F'} {A_{F,F'}^{{rot}}Im\left[ {{\cal V}\left( {\nu  - {\nu _{F,F'}}} \right)} \right]} $. In our experiments, the optical rotation $\varphi$ is small, so the assumption $\sin \left( {2\varphi } \right) \approx 2\varphi $ has been employed in Eq. \ref{sig}. However, in the case where $\varphi$
is relatively large, a more complete description without this approximation is also available for numerically optimizing the parameters of EPMx AMs.

\subsection{Light intensity and ellipticity analysis}

\begin{figure}[htp]
	\centering\includegraphics[width=8cm]{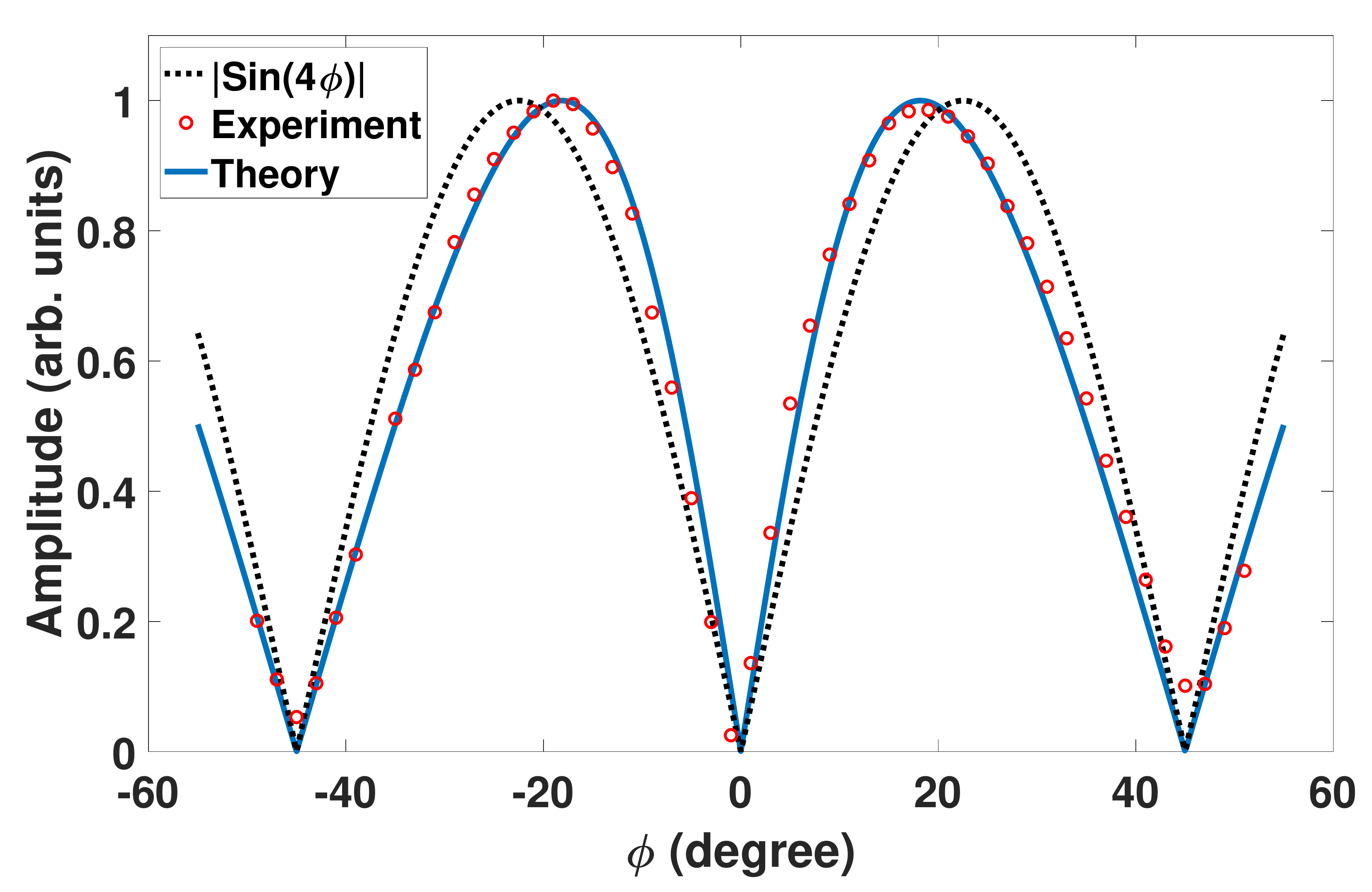}
	\caption{The signal amplitude as a function of the $\lambda/4$ waveplate angle $\phi$. Red hollow dots represent experimentally measured results. The blue solid line is the theoretical prediction taking ellipticity variations into account. The dashed line shows the function of $\left| {\sin \left( {4\phi } \right)} \right|$.}
	\label{ellipticity}
\end{figure}

Before revealing the optical frequency response of the EPMx AM signal, the influences of the optical absorption to the light intensity as well as its ellipticity need to be investigated. The incident photon fluxes of $\sigma^+$ and $\sigma^-$ components are $\frac{{{\rm{1 + }}s}}{2}{\Phi _0}$ and  $\frac{{{\rm{1 - }}s}}{2}{\Phi _0}$, respectively. In the cell, polarized atoms present different absorption rates to $\sigma^+$ and $\sigma^-$ photons. Assuming that the polarization of atoms is spatially uniform, the emergent photon fluxes of $\sigma^+$ and $\sigma^-$ parts decay to
\begin{equation}
\begin{split}
\Phi _1^ +  &= \frac{{{\rm{1 + }}s}}{2}{\Phi _0}{e^{ - nl{\sigma _{total}}\left( {1 - p} \right)}},  \\
\Phi _1^ -  &= \frac{{{\rm{1 - }}s}}{2}{\Phi _0}{e^{ - nl{\sigma _{total}}\left( {1 + p} \right)}}.
\end{split}
\label{photonflux}
\end{equation}
Then we can get the photon flux and ellipticity of the emergent light as ${\Phi _1} = \Phi _1^ +  + \Phi _1^ -$ and ${s_1} = {{\left( {\Phi _1^ +  - \Phi _1^ - } \right)} \mathord{\left/
		{\vphantom {{\left( {\Phi _1^ +  - \Phi _1^ - } \right)} {\left( {\Phi _1^ +  + \Phi _1^ - } \right)}}} \right.
		\kern-\nulldelimiterspace} {\left( {\Phi _1^ +  + \Phi _1^ - } \right)}}$.
Therefore the light absorbing medium, i.e., the polarized alkali-metal atomic ensemble, makes a change in the light ellipticity. The variation is more obvious in resonant or near-resonant condition due to bigger absorption cross-section. Besides, greater atomic number density and longer cell length will exacerbate this effect. It is easy to find that $1/{\sqrt 2 }$ is the optimal value of $s$ by maximizing $s\sqrt {1 - {s^2}} $, provided we treat ellipticity as a constant. The corresponding $\lambda/4$ waveplate angle $\phi=\pi/8$ has become a standard configuration when we employ elliptically polarized laser in AMs \cite{Shah09, Ding06}. However we find that a slight deviation of $\phi$ from $\pi/8$ facilitate a better performance of AMs, especially in the near-resonant condition in our case. The experimentally measured amplitude of the photoelectric signal of our EPMx AM, at low light intensity of 150 uW and temperature of 60 $^{\circ}$C, is shown in Fig. \ref{ellipticity} as a function of $\phi$. A mismatch from the behavior as $\left| {\sin \left( {4\phi } \right)} \right|$ is confirmed, while introducing the change of ellipticity makes the theoretical description more consistent with experimental results. In this case, the effect of ellipticity variation reaches a considerable extent and the
optimal angle $\phi_{opt}={18.2^ \circ }$ can be determined, corresponding to an initial ellipticity of $s \approx 0.59$.

\subsection{Laser frequency optimization}

\begin{figure}[htp]
	\centering\includegraphics[width=8cm]{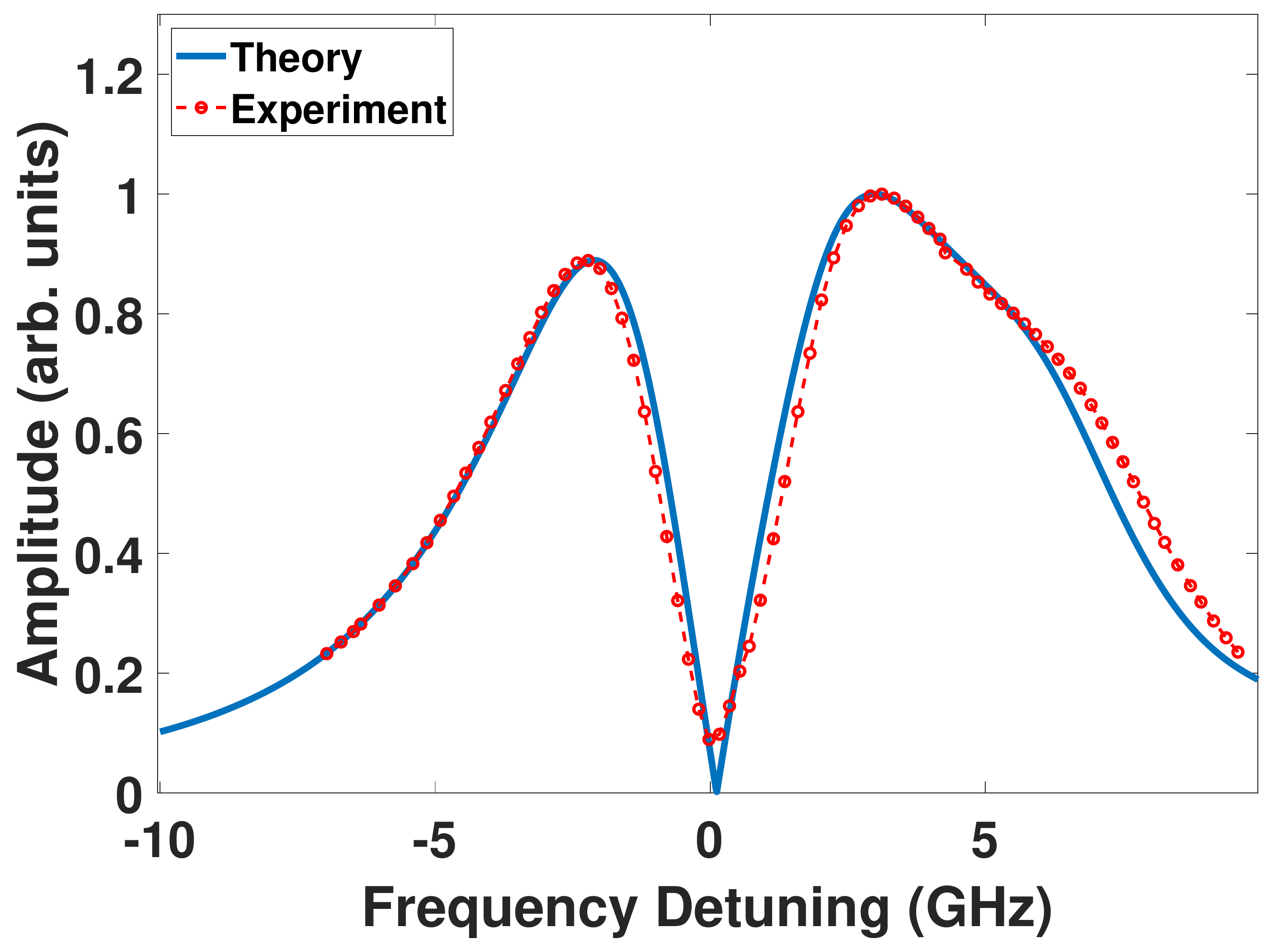}\caption{The signal amplitude as a function of  frequency detuning from D$_1$ $F=2$ $\to$ $F'=1$ transition. Red dots represent experimental data. The blue line is theoretical prediction as Eq. (\ref{sig}). The experiment is taken at temperature of 60 $^{\circ}$C and light intensity of 0.83 $\rm{\mu}$W/$\rm{mm^2}$.}
	\label{frequency}
\end{figure}

Having modified the parameters of light intensity and ellipticity, now we turn to an investigation on the laser frequency optimization. As shown in Eq. (\ref{sig}), the laser frequency response lineshape of EPMx AM signal strongly depends on  $\sigma_{\rm{total}}$ and $C_{\rm{rot}}$, which closely relate to the pumping and detecting processes, respectively. We can not choose the resonant frequency like the conventional $M_x$ magnetometer, since the value of $C_{\rm{rot}}$ is too close to zero at this frequency, although it allows a strongest absorption cross-section $\sigma_{\rm{total}}$. Far off-resonant frequencies which are usually employed in conventional OR detection mode or high atomic density condition \cite{Shah09} are also not suitable due to a significant reduction of pump rate. Not only that but far off-resonant frequencies are adverse to obtaining a big refraction factor $C_{\rm{rot}}$. To seek a optimal point in the near-resonant region, we measured the amplitudes of the photoelectric signals as scanning the laser frequency. A frequency-stabilized laser was used to form a beat frequency system with the concerned laser to measure the amount of detuning. The experimental data are shown in Fig. \ref{frequency} as a function of the frequency deviation from the transition $F=2$ $\to$ $F'=1$ of $^{87}$Rb D$_1$ line. The experimental results are well consistent with the theoretical frequency response curve depicted by Eq. (\ref{sig}). Both theoretical and experimental results point to the blue shift of 2-4 GHz as an optimal operating range for an EPMx AM. 

\section{Experimental results and discussion}

Choosing the phase signal for following studies, we can characterize the sensitivity of the magnetometer in terms of the noise equivalent magnetic flux density $\delta$B \cite{Groeger06}, expressed as 
\begin{equation}
\text{$\delta $B}=\frac{ V_n  \Delta \omega _{\text{HW}}^\theta }{ \gamma  k}.
\label{sensitivity}
\end{equation}
Here $V_n$ is the noise level charactered by the square root of the power spectral density (PSD) of the phase output of lock-in amplifier in resonant condition. We estimate $V_n$ as the average noise level between 1 and 10 Hz. Although it's called  ``noise level'', $V_n$
is actually closely determined by the signal-to-noise ratio of the magnetometer signal. $k=\frac{9}{\pi }$ V$/$rad is the phase scale factor of lock-in amplifier, by which we can convert the voltage output to phase representation. Therefore the noise level $V_n$ and the half resonance width $\Delta \omega _{\text{HW}}^\theta$ are two essential indicators for the sensitivity analysis of an EPMx AM and its comparison with the CPMx AM.

\begin{figure}[htp]
	\centering\includegraphics[width=10cm]{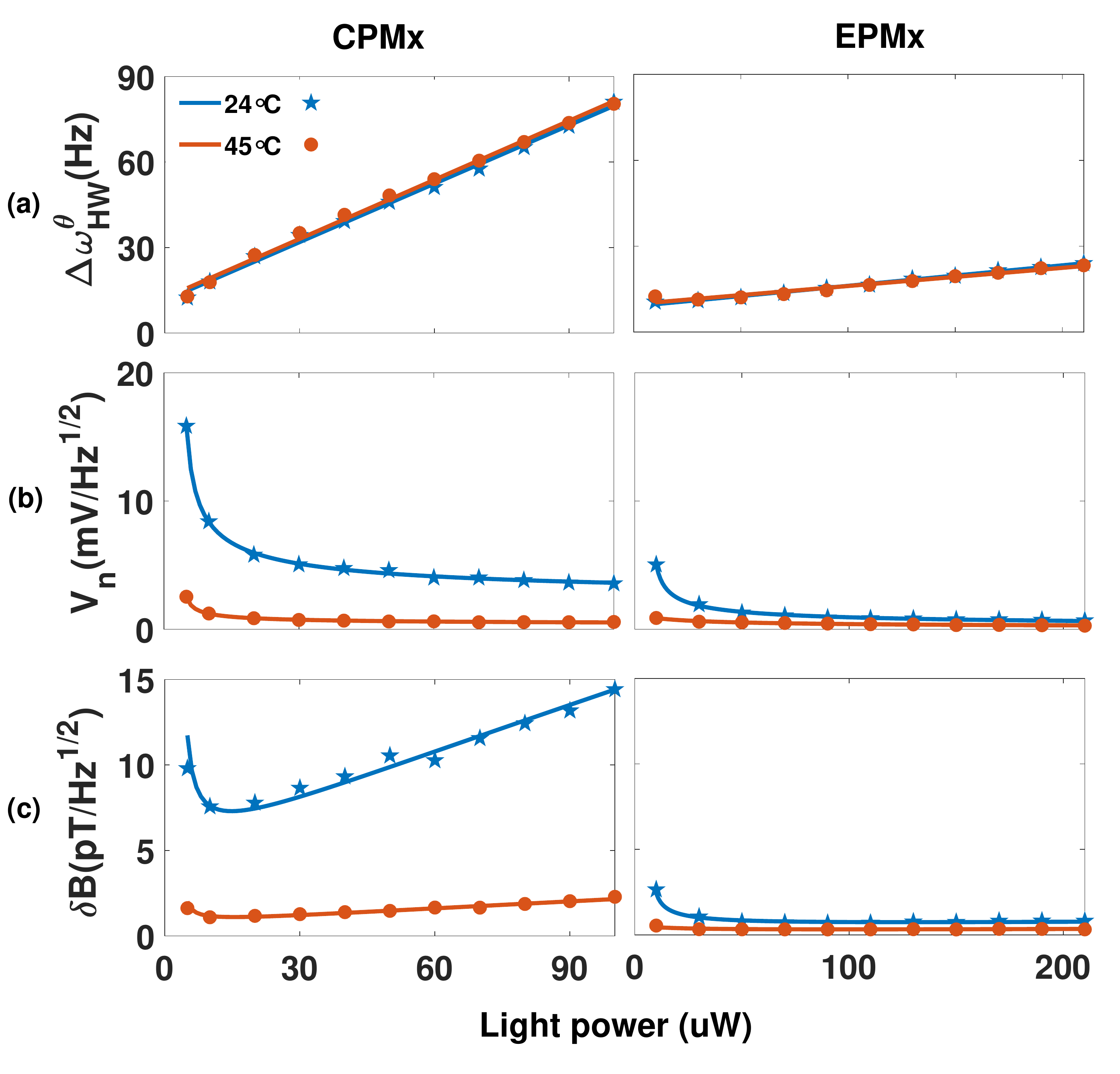}
	\caption{Sensitivity characterizations of CPMx and EPMx AMs as a function of incident light power at temperatures of 24 $^{\circ}$C and 45 $^{\circ}$C. Red dots and blue pentacles represent the measured results at 24 $^{\circ}$C and 45 $^{\circ}$C, respectively.
		(a) The half resonance width. The solid lines are linear fitting results. Note that two curves of each AM configuration are almost overlapped. (b) The noise level. The solid lines are fitting results with the function $1/\sqrt{{\cal {I}}_{in}}$. (c) The noise equivalent magnetic flux density. The solid lines are obtained by substituting  $V_n$ and $\Delta \omega _{\text{HW}}^\theta$ of Eq. \ref{sensitivity} with the fitting results of (a) and (b).}
	\label{noise}	
\end{figure}

Besides the laser frequency and ellipticity, laser intensity is another important parameter greatly influencing the performance of AMs. A stronger laser usually means a more intense signal, while it also results in a wider resonance linewidth. For the sake of a comparison between the EPMx AM and  its conventional counterparts at respective optimal conditions, we measured the resonance linewidths and noise level of both configurations as varying the incident light intensity. The results are shown in Fig. \ref{noise}. As expected, the resonance linewidth linearly dependents on the light intensity due to the linearly growing pumping rate. 
It is readily comprehensible that the linewidth of EPMx AM is narrower than CPMx AM at the same laser power on account of a laser frequency detuning of several gigahertzes and a small ellipticity in the EPMx AM. However, at respective optimal operating points (10 uW for CPMx AM and 90 uW for EPMx AM, Fig. \ref{noise}c), there is no significant difference in linewidths between these two AMs. We also conducted the experiments at temperatures of 24 $^{\circ}$C and 45 $^{\circ}$C, respectively. As main mechanisms contributing to the intrinsic relaxation rate, the rates of spin-exchange collisions and spin-destruction collisions are little affected in this temperature region. Therefore, we observed almost coincident linewidth curves at 24 $^{\circ}$C and 45 $^{\circ}$C. An intrinsic linewidth of about 10 Hz can be inferred at zero light power from the measured results. From Fig. \ref{noise}b, we can see that $V_n$ decreases when the light power ${\cal {I}}_{in}$ increases. The decay behavior can be well described as the tendency $1/\sqrt{{\cal {I}}_{in}}$. The EPMx AM shows its advantage in noise level by greatly suppressing common mode noise with OR detection mode. The suppression effect is particularly pronounced in the case where the atomic number density is low, or at a low temperature. At high temperature, more atoms are involved into interaction with the light, pushing the signal-to-noise ratio to the limit. This is why the CPMx shows a comparable $v_n$ at high temperature. 

\begin{figure}[htp]
	\centering\includegraphics[width=10cm]{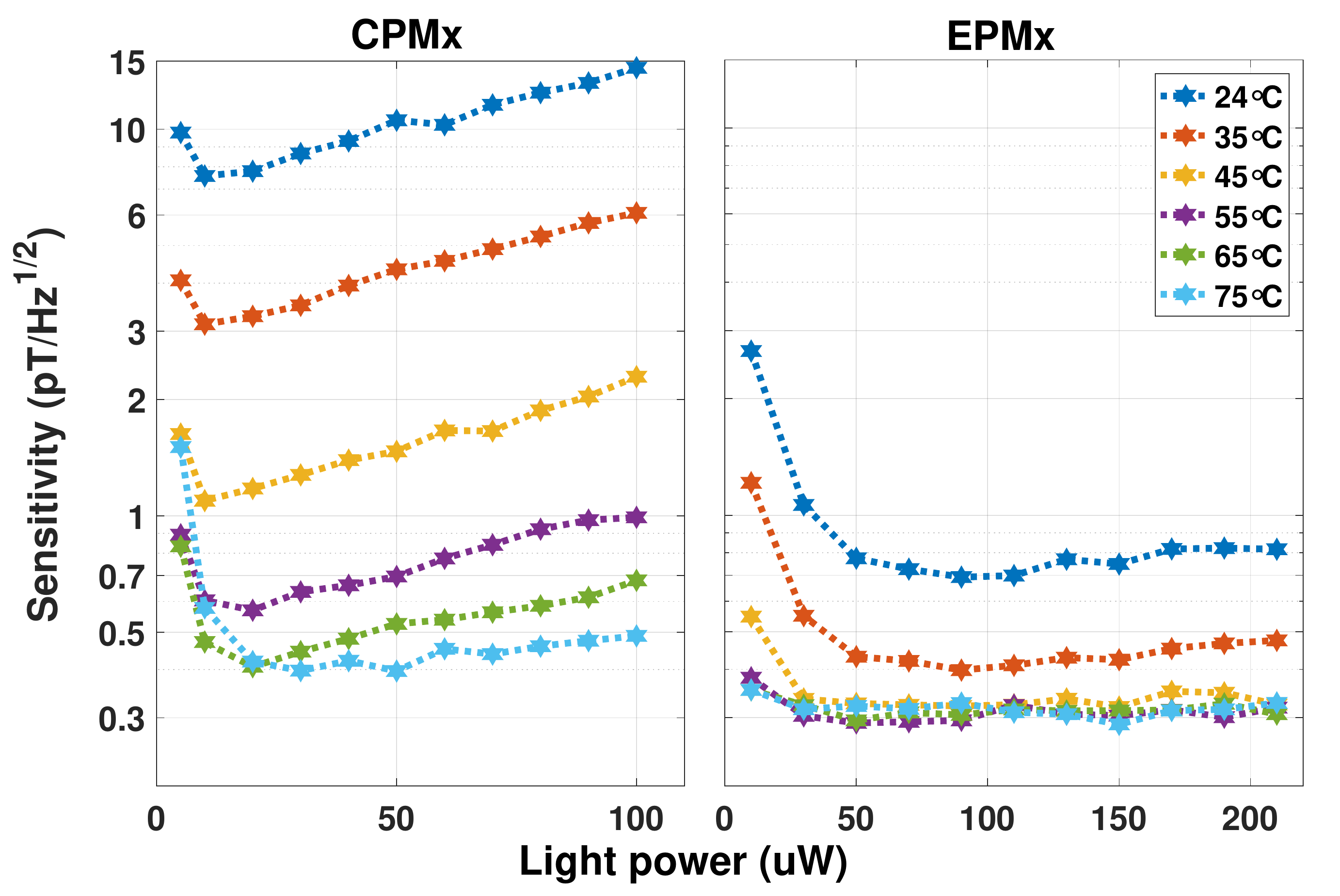}
	\caption{Sensitivity comparison between CPMx and EPMx AMs as varying the incident light power and temperature in the range 24 $^{\circ}$C to 75 $^{\circ}$C.}
	\label{sensitivity}	
\end{figure}

At different temperatures from 24 $^{\circ}$C to 75 $^{\circ}$C, we present the sensitivities of CPMx AM and EPMx AM as a function of the light power in Fig. \ref{sensitivity}. The light power affect CPMx AM more severely, while it has little effect on EPMx AM over 50 uW. At 24 $^{\circ}$C and respective optimal light powers, we obtained a sensitivity of 0.69 $\rm pT/\sqrt{Hz}$ with EPMx AM, which is an order of magnitude improvement than CPMx AM. The optimal sensitivity increases as the temperature rises. At 45 $^{\circ}$C, the sensitivity of EPMx AM is 0.32 $\rm pT/\sqrt{Hz}$, while it's 1 $\rm pT/\sqrt{Hz}$ for CPMx AM. When the temperature is higher than 45 $^{\circ}$C, we won't get much in return by increasing the temperature. The sensitivity is still around 0.3 $\rm pT/\sqrt{Hz}$ for EPMx AM even at 75 $^{\circ}$C, while the sensitivity of CPMx AM approaches 0.4 $\rm pT/\sqrt{Hz}$ at this temperature. 


\section{Conclusions}
In conclusion, we have presented and characterized a $M_x$ magnetometer with a single near-resonant beam of elliptically polarized light based on $^{87}$Rb atoms. Theoretical analyses have been carried to clarify main mechanisms affecting the performance of an EPMx AM. Taking the light intensity and ellipticity variations into account, we present the output signal of EPMx AM in the analytical form. Moreover, main theoretical results in this paper were experimentally verified. Based on these theoretical and  experimental results, we optimized the important optical parameters, such as light ellipticity, frequency and power. A great portion of effort has been contributed to the comparison between CPMx AM and EPMx AM. At respective optimal conditions, EPMx AM shows a substantial improvement in sensitivity. Especially at relatively low temperatures, where the atomic number density is low,  more than one order of sensitivity improvement has been obtained. To be specific, at room temperature of 24 $^{\circ}$C, a sensitivity of 0.69 $\rm pT/\sqrt{Hz}$ has been achieved with EPMx AM, much better than 7.6 $\rm pT/\sqrt{Hz}$ of CPMx AM working at its optimal condition. We found that it's not necessary to continue to raise the temperature of our EPMx AM after 45 $^{\circ}$C, since there will be no significant improvement than 0.32 $\rm pT/\sqrt{Hz}$ obtained at 45 $^{\circ}$C. Finally, we optimized the sensitivities of CPMx AM and EPMx AM to 0.4 $\rm pT/\sqrt{Hz}$ and 0.29 $\rm pT/\sqrt{Hz}$ at 75 $^{\circ}$C, respectively.  

The results in this paper show that introducing elliptically polarized laser to $M_x$ magnetometer is of great significance. It not only preserves the compact potential with one single beam configuration, but also allows a great sensitivity improvement. Excellent performance extends the application range of $M_x$ AM with uncoated cell in unheated environment. As an example, we have designed a compact probe of EPMx AM and 
realized high-quality magnetocardiography measurements at low temperature. The EPMx AM is particularly suitable for practical applications of magnetometer array. Using uncoated cells can eliminate the inconsistency caused by coating process and reduce the cost. Working at room temperature avoids the structural complexity caused by heating and insulation units. The EPMx AM also has low power consumption characteristic, which is important in some practical applications, such as long-term outdoor geomagnetic measurements and wearable magnetocardiography measurements.

\section{ACKNOWLEDGMENTS}
This work are supported by the National Key Research and Development Program of China (Grant No. 2017YFC0601602) and the National Natural Science Foundation of China (Grant No. 11605153, 61727821, 61475139).

\bibliographystyle{unsrt}
\bibliography{references}

\end{document}